\documentclass[conference]{IEEEtran}
\usepackage{cite}
\usepackage{amsmath,amssymb,amsfonts}
\usepackage{algorithmic}
\usepackage{graphicx}
\usepackage{textcomp}
\usepackage{xcolor}
\usepackage{comment}
\usepackage{algorithm}
\usepackage{subfigure}
\usepackage{color}
\usepackage{xcolor}
\makeatletter
\newcommand*{\rom}[1]{\expandafter\@slowromancap\romannumeral #1@}
\makeatother
\usepackage{setspace}
\newcommand{\subheader}[1]{ \textbf{#1}}

\usepackage[normalem]{ulem}

\usepackage{tikz}

\ifCLASSINFOpdf
\else
\fi
\begin{document}

\title{A Preliminary Study of Neural Network-based Approximation for HPC Applications}

\author{\IEEEauthorblockN{Wenqian Dong}
\IEEEauthorblockA{EECS, UC Merced \\
wdong5@ucmerced.edu}
\and
\IEEEauthorblockN{Luanzheng
 Guo}
\IEEEauthorblockA{EECS, UC Merced \\
lguo4@ucmerced.edu}
\and
\IEEEauthorblockN{Dong Li}
\IEEEauthorblockA{EECS, UC Merced \\
dli35@ucmerced.edu}

}

\maketitle

\pagestyle{plain}


\begin{abstract}
Training neural network often uses a machine learning framework such as TensorFlow and Caffe2. These frameworks employ a dataflow model where the NN training is modeled as a directed graph composed of a set of nodes. 
Operations in neural network training are typically implemented by the frameworks as primitives and represented as nodes in the dataflow graph. Training NN models in a dataflow-based machine learning framework involves a large number of fine-grained operations. Those operations have diverse memory access patterns and computation intensity. How to manage and schedule those operations is challenging, because we have to decide the number of threads to run each operation (concurrency control) and schedule those operations for good hardware utilization and system throughput.

In this paper, we extend an existing runtime system (the TensorFlow
runtime) to enable automatic concurrency control and scheduling of operations. 
We explore performance modeling to predict the performance of operations with various thread-level parallelism. Our performance
model is highly accurate and lightweight. Leveraging the performance model, our runtime system employs a set of scheduling strategies that co-run operations to improve hardware utilization and system throughput. 
Our runtime system demonstrates a big performance benefit. Comparing with using the recommended configurations for concurrency control and operation scheduling in TensorFlow, our approach achieves 33\% performance (execution time) improvement on average (up to 49\%) for three neural network models, and achieves high performance closing to the optimal one manually obtained by the user.
\end{abstract}

%
\IEEEpeerreviewmaketitle
\section{Introduction}
\label{sec:intro}
Large-scale scientific simulations drive scientific discovery across many domains.
Those scientific simulations increasingly face performance problems, because of hardware heterogeneity, deep memory hierarchy, and massive thread-level parallelism. Addressing those problems often requires domain scientists to use sophisticated compiler and runtime techniques to optimize HPC programs. However, domain scientists are often not skillful computer scientists and may find program optimization time-consuming and daunting. In this project, we study how to use an alternative approach, machine learning, to effectively improve the performance of scientific simulations without losing simulation quality. 

Machine learning, as a tool to learn and model complicated (non)linear relationships between input and output data sets, has shown preliminary success in some HPC problems. Using machine learning, scientists are able to augment existing simulations
by improving accuracy and significantly reducing latencies. 
For example, scientists working to detect neutrinos at Fermi National lab have realized a 33\% improvement in neutrinos detection using a convolutional neural network~\cite{ichep16:radovic};
Scientists achieve Bose-Einstein Condensates state in only 10-12 experiments using machine learning instead of 140 experiences using traditional models, which reduces the simulation time by 10 times~\cite{sr15:widley}. 
Other successful examples of using machine learning for HPC include recognizing extreme weather events in large-scale climate simulations at Lawrence Berkeley National Lab (LBNL)~\cite{nips17:racha}, and precision medicine for cancer~\cite{candle_anl} at Argonne National Lab. 

However, our current methodology to apply machine learning to scientific simulations has limitations. In particular, the current methodology is rather ad-hoc and application-specific. There is no systematic and principled approach to enable general application of machine learning methods to scientific simulations. 
The lack of a systematic and principled approach
is especially problematic to ensure high simulation quality when using machine learning.  Furthermore, using machine learning requires domain scientists to have machine learning knowledge, while they usually do not have sufficient machine learning background. How to make machine learning techniques widely accessible and usable to domain scientists is largely unexplored. 


Our ongoing research work is to create a general framework to apply neural network-based models to HPC applications. 
In particular, we want to use the neural network to approximate and replace code regions within the application to improve performance (i.e., reducing the execution time) of the application. 
In this paper, we present our preliminary study and results.

To use a neural network model to replace a code region, we face multiple research challenges in our preliminary work. 
First, we must make sure that our neural network can bring a performance benefit.
This means the execution time (inference time) of the neural network should be shorter than that of the replaced code region. 
To address this issue, we use a trial-and-error method to try different neural network models based on the performance of the original code to decide which model should be used.
Using such a performance-driven approach to select the model separates us from the existing work where the model accuracy is often used to select the model.

Secondly, we must determine the appropriate input and output variables for the neural network models.
Those variables are also the input and output variables of the replaced code region. 
We define the input and output variables of a code region based on the memory access pattern (i.e., the read/write) and variable liveness analysis.

We make the following contributions in this paper:

\begin{itemize}
    \item We explore the feasibility of using neural networks to approximate certain computation in HPC applications. Using the Newton-Raphson method and L-J potential in LAMMPS (a molecular dynamics simulation code) as examples, we show up 2.7x and 2.46x speedup, respectively. 
    
    \item We study performance (execution time) implications of using different neural network models on the HPC applications. We also study the impact of using different neural network models on the approximation accuracy.
    
    \item We introduce a general and preliminary workflow to identify code regions and apply neural network models to replace them.
\end{itemize}

As a preliminary work, for our study of the L-I potential, we have not considered the impact of using neural networks on the application result correctness (we consider so for the Newton-Raphson method). We hope to extend our study in the future work.

\section{Background}
\label{sec:bg}
We reveal the relevant background information in this section. 

\subsection{Machine Learning-based Approximation}
Machine learning-based approximation is in essence approximate computing.
We study machine learning-based approximation rather than other approximate computing techniques because it has big advantages over other approximate computing techniques.
(1) Other techniques, such as loop perforation~\cite{sidiroglou2011managing, 5470469}, random task discarding~\cite{Rinard:2006:PAB:1183401.1183447}, and synchronization relaxation~\cite{Rinard:2007:UEP:1297027.1297055, Campanoni:2015:HRP:2738600.2738630}, have specific requirements on the code structure to be approximated. The requirement could be a loop structure, a task-based execution model, or communication synchronization. Machine learning-based approximation does not have such a constraint on code structures. (2) Other techniques cannot provide portability on heterogeneous hardware as machine learning-based approximation. 
(3) Other techniques cannot provide flexible quality control as machine learning-based approximation. 
Quality control means controlling the output quality of the replaced code region. 
Having flexibility for quality control gives us a large room to explore the tradeoff between performance and accuracy. 
The quality control in other techniques is typically constrained by the code structure to implement approximation (e.g., the number of iterations in a loop structure for loop perforation), while machine learning does not have such constraint. 
We can use different machine learning models with different configurations to provide a variety of output quality with different performance.

\subsection{Two Applications for Study}
We study two HPC applications, which are an implementation of the Newton-Raphson method and LAMMPS. 

\textbf{Newton-Raphson method.}
Newton-Raphson method is a root-finding algorithm to successively search for a better approximation of the roots to a real-valued function. 
To achieve the goal, the Newton-Raphson method repetitively uses a derivation to 
to find the best way to approach the optimal solution. 
Assuming a root for a function $f(x)$ is needed (the function $f(x)$ is defined over the real numbers $x$) and the function $f(x)$ satisfies the assumptions made in the derivation of the formula, i.e., $f'(x)$, 
the function $f(x)$ can be expressed as follows according to the Taylor series.

\begin{equation}
\label{eq:newton_1}
 f(x)= f(x_0)+f'(x)(x-x_0).
\end{equation}

To search an approximate root of the equation, an initial solution $x0$ (a random real-value or a value defined by users) is used as the first step to find the root of the function $f(x)$.
After that, we repeatedly use Equation~\ref{eq:newton_3}
to find a new solution. 

\begin{equation}
\label{eq:newton_3}
 x_{n+1}= x_n-\frac{f(x_n)}{f'(x_n)},
\end{equation}

The above iterative process continues until  
a termination condition is satisfied.
The termination condition is defined as
$|x_{n+1}-x_n|<\epsilon$ or $|f(x_{n+1})|<\delta$, where $\epsilon$
or $\delta$ are defined by the user.

\textbf{The Lennard-Jones (LJ) potential in LAMMPS.}
LAMMPS~\cite{plimpton2007lammps} is a molecular dynamic simulation tool developed by Sandia National Laboratory. 
LAMMPS can be used for modeling particles movement at multiple scales (atomic, meso, continuum scales) in parallel. 
In LAMMPS, we often calculate a ``potential function'' for a pair of atoms. 
The L-J potential is a potential function in LAMMPS. It is a simple model to approximate the force interaction between a pair of neutral atoms or molecules. 
Equation~\ref{equation:lj} shows the computation of the L-J potential. 

\begin{small}
\begin{equation}
\label{equation:lj}
U(\overrightarrow{R})=\\
\left\{
\begin{array}{rcl}
\sum\limits_{i}\sum\limits_{j\neq i}4\epsilon_{ij}
\left[\left(\frac{\sigma_{ij}}{r_{ij}}\right)^{12}-\left(\frac{\sigma_{ij}}{r_{ij}}\right)^{6}\right], & r_{ij}<r_{c}=2^{1/6}\sigma\\
0, & r_{ij} \geqslant r_{c}
\end{array}
\right.
\end{equation}
\end{small}

In Equation~\ref{equation:lj}, for a pair of atoms $i$ and $j$ located at $\overrightarrow{r_{i}}$ and $\overrightarrow{r_{j}}$, we have $\overrightarrow{r_{ij}} = \overrightarrow{r_{i}} - \overrightarrow{r_{j}}$ and $r_{ij} = |\overrightarrow{r_{ij}}|$. In Equation~\ref{equation:lj}, the parameter $\epsilon$ governs the strength of atom interaction, and the parameter $\sigma$ defines the length scale.

\subsection{Neural Network}
The existing work shows that
the neural network can be used to approximate code regions for applications in diverse domains~\cite{micro2012neural,Esmaeilzadeh:2012:ASD:2150976.2151008,carleo2017solving,ml_fft,md_ml17}.  
In this paper, we also use the neural network to replace computation- or memory-intensive code regions. We briefly review the neural network as follows.

Neural network(NN) is a popular machine learning model and it broadly includes CNN (convolutional neural network), RNN (recurrent neural network), and GAN (generative adversarial network). 
A neural network is composed of nodes (i.e., neurons) and edges. 
Nodes are organized into layers in the neural network; nodes across layers are connected by edges; each edge has a weight. We learn the weights when training the neural network. 
Typically, there are three types of layers: the input layer, the hidden layer, and the output layer. 
A node of an input layer is some input data of the neural network; a node of a hidden/output layer is the weighted sum of the input data of the node; an output layer can have one or more nodes: it depends on whether the neural network is for a classification problem or a regression problem. 
In addition, there are bias nodes at each layer, which are used for compromising noise in the input data to avoid overfitting. 
There can be an activation function such as sigmoid or rectifier for a layer to rectify the incoming data for the next layer. 

In our work, we use the supervised learning to replace computation in HPC applications with neural networks such as CNN. 

\section{Problem Definition}
\label{pro_def}
This paper particularly targets the following research problem.
We characterize HPC applications as a set of code regions.
A code region is simply a block of code. It can be a loop structure; it can also be a function. 
 

We selectively replace code regions with neural networks to improve performance (i.e., shortening execution time) of HPC applications.
The neural network should use the same input and output variables as the original code region.

We choose code regions to replace, based on two criteria: (1) The code region must be time-consuming and its execution time takes a large portion of the total execution time of the application. 
(2) Replacing the code region should not impact the correctness of the application outcome. This indicates that the application itself should be able to tolerate the approximation introduced by the replacement of the code region. 




\textbf{Time-consuming code regions.}
According to Amdahl's law, we can achieve the theoretical maximum speedup
by improving the performance of the most time-consuming portion of a workload.
In our work, we claim a code region is time-consuming, if a single invocation of the code region takes a large portion of the total execution time of the application, or the code region is repeatedly executed and the accumulated execution time of the code region takes a large portion of total execution time of the application.


To select a code region to replace, besides measuring its execution time, we particularly pay attention to controlling flows in the code region. 
The control flows can prevent compiler optimization and effective instruction scheduling, hence causing performance loss. Furthermore, When running the code region on a SIMD architecture such as GPU, the control flows can cause idling threads and decrease hardware utilization. Hence, we want to replace such a code region with a neural network, such that we can remove control flows within the code region.

\textbf{Approximability of HPC applications.}
Many HPC applications can tolerate computation inaccuracy caused by approximate computation. This has been demonstrated in the existing work~\cite{khudia2015rumba, Krishnan:2016:IDA:2872427.2883026, asplos14:samadi, Zhang:2015:AAC:2840819.2840934}. 
In fact, HPC applications themselves are approximate in nature. 
For example, the molecular dynamic simulation only models the force between atoms that are close enough in the physical space. The long-distance force is just ignored. Hence, even without introducing machine learning-based approximation, HPC applications already have some approximation.  

Furthermore, many HPC applications have a threshold to determine when the final application outcome is acceptable or when the simulation should be terminated. 
Such a threshold-based approach allows the HPC applications to tolerate approximate computation. 

In our study, we assume that HPC applications have explicit requirements on the final simulation quality (e.g., a threshold) to ensure approximation correctness.  
Using neural networks to replace code regions can generate computation inaccuracy in the middle of scientific simulations (i.e., HPC applications), but the final simulation result must meet the requirements of domain scientists on the final simulation quality.

\textbf{Input and output variables of code regions.}
Given a code region, we classify the variables within the code region as input variables, output variables, and internal variables. 
Input variables are those that are declared outside of the code region and referenced in the code region. Output variables are those that are written in the code region and read after the code region. 
Other variables that the region writes to or reads from are internal variables. 
A code region can be executed many times during the application execution. 



\textbf{Concerns on training time.}
A neural network must be trained before it is deployed in an HPC application to replace a code region. 
When evaluating performance benefit of the neural network, the training time must be considered.
A trained neural network is expected to give a prediction of the values of output variables. 
Note that if an input or output variable changes its size (e.g., an input 2D matrix changes its size from 512x512 to 1024x1024), then the neural network must be re-trained.
Hence, the replaced code region should have 
fix-sized input and output variables, and 
must be repeatedly executed to have sufficient performance benefit, such that 
we can avoid repeatedly training the model and the overhead
of modeling, training is amortized and justified. 

A large number of scientific simulation applications have
code regions that meet the above requirements.
For those code regions, we only need to train the neural network once. Training time should be less than the performance benefits of using the neural network-based approximation. We give two example cases as follows.

\begin{itemize}
\item The Lattice Boltzmann method (LBM) has been widely employed in computational fluid dynamics with broad applications (e.g., multiphase flows, reacting flows, phase-change heat transfer, complex flows in porous media, simulations of microfluidics and nanofluids). 
In the parallel implementation of LBM, the computational domain is divided into subdomains with fixed-size input and out variables. The sizes of those variables are independent of the input problem size of LBM method. 

\item Climate modeling (including atmosphere modeling CAM, ocean modeling POP2, land surface modeling CLM4 and sea ice modeling CICE4~\cite{cesm_models}) can take a very long time (weeks or even months) to solve different equations. During the model simulation, many code regions have fixed-size input and output variables, and repeatedly executed.
\end{itemize}


In this paper, we choose the Newton-Raphson method and L-J potential in LAMMPS as our targets to replace, because they are very frequently used. The performance benefit of replacing them can easily overweight the training time.

\section{Neural Network-based Approximation}
We discuss how to use the neural network to approximate code regions in this section.

\subsection{General Methodology}
Given an application, we use gprof to identify the most time-consuming functions. Those functions are candidate code regions to be replaced. In our study, we replace the whole Newton-Raphson method, because its implementation is simple enough to be treated as a function to replace. 
We replace the computation of the L-J potential in LAMMPS because it is simple enough for our preliminary study. Also, the L-J potential is the most time-consuming computation for some input problems of LAMMPS (we use in.lj.5 as the input problem of LAMMPS).

After code regions are selected, we need to determine the input and output variables of the code region. This can be done based on compiler analysis. 


After the input and output variables of the code region are decided, we need to build a neural network. 
Building a neural network involves a determination of the network topology (e.g., how many layers and how many neurons in each layer and what are the activation functions in the neural network).
Furthermore, there are various types of neural network, such as Convolutional Neural Network (CNN) and Recurrent neural network (RNN). We need to decide which type of neural work should be chosen.
From the performance perspective (execution time), we decide the network topology and which type of neural network should be chosen based on the computation complexity of each candidate neural network. The execution time (i.e., inference time) of the neural network should be shorter than the replaced code region.


We extract the code region out of the application as a standalone application.
Then we randomly generate input data, feed them into the code region, and then collect output data. Each pair of input and output data is a training example.
Note that when we generate random input data, the input data must meet the requirement of the application on the input data.  

We describe how to replace the Newtwo-Raphson method and the computation of the L-J potential in LAMMPS as follows.

\subsection{Newton-Raphson method}
The Newton-Raphson method is widely used in finding an approximation root of an equation. 
Algorithm~\ref{alg:Newton-Raphson method} generally depicts the Newton-Raphson method. The Newton-Raphson method can be time-consuming because it iteratively uses
a derivation to find the best way to approach the optimal solution.
Sometimes the Newton-Raphson method has to  
use a large number of iterations to find a good solution.
To determine if a solution is good, the Newton-Raphson method
examines if the difference between the current solution and the immediately last solution is smaller than a threshold (see Line 3 in Algorithm 1).
Such a threshold-based approach allows the Newton-Raphson method to tolerate
approximate computation. 

\begin{algorithm}[h]
\caption{\small Newton-Raphson method}
\label{alg:Newton-Raphson method}
\begin{algorithmic}[1]
\REQUIRE {Function $f$; Truncate error $E$; Initial assumption $x_{0}$; default number of iterations $N$.}
\ENSURE {Final solution $F$}
\STATE $f(x)=0$;
\STATE $x=x_0$;
\IF{$|x_i-x_{i-1}<E|$ and $i < N $}
    \FOR{each assumption $x_{i}$}
        \IF{ $f(x_i)$ is differentiable}
          \STATE $x_{i+1}=x_i-\frac{f(x_i)}{f'(x_i)}$
          \STATE $i=i+1$
        \ENDIF
    \ENDFOR
\ENDIF
\RETURN $x_{n}$

\end{algorithmic}
\end{algorithm}

\begin{figure}[!t]
  
    \centering
    \includegraphics[width=0.36\textwidth, height=0.20\textheight]{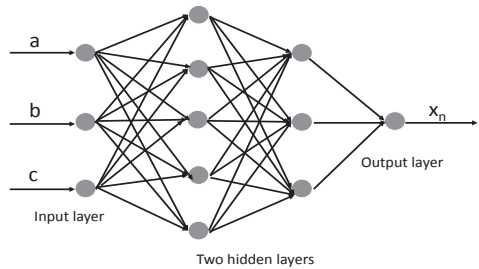}
       \vspace{-15pt}
        \caption{A 4-layers DNN model.} 
  \vspace{-5pt}
  \label{fig:a small model}
\end{figure}

We can easily identify the input and output variables of the Newton-Raphson method. 
Assuming that the target equation for the Newton-Raphson method to solve is a quadratic function $f(x)=ax^{2}+bx+c$, then the input variables are $a, b$ and $c$,  and the output variable is the final solution $x_{n}$.


we use a fully-connected NN model to replace the whole Newton-Raphson code in our study.
Our model is shown in Figure~\ref{fig:a small model} consists of fully connected neurons. Those neurons are organized as an input layer, two hidden layers, and an output layer.
The input of the model is three variables and the output of the model is a single variable. We use a $3\times5\times3\times1$ NN and use the Momentum backpropagation approach when training the neural network.

We use $102,400$ samples for model training and $3072$ samples for model validation.
Using this 4-layers fully connected neural network, we achieve good modeling accuracy, $89.47\%$ after $5,000$ training steps.  

\begin{figure}[!t]
    \centering
    \includegraphics[width=0.50\textwidth, height=0.40\textheight]{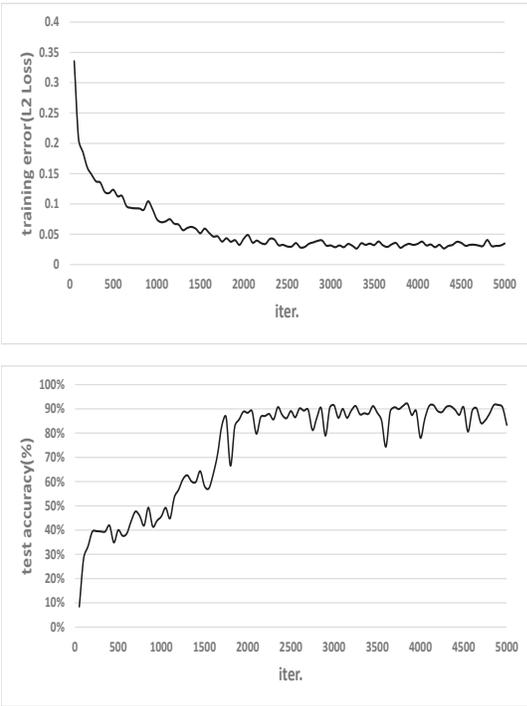}
       \vspace{-15pt}
        \caption{The training loss and test accuracy across time steps.} 
  \vspace{-35pt}
    \label{fig:Newton_l2loss}
\end{figure}

Figure~\ref{fig:Newton_l2loss} shows how the training error and test accuracy vary as we increase the number of training steps. 
The training error is measured by L2 loss (squared error). 
From the figure, we can see the convergence of the L2 loss and prediction accuracy after 5,000 training steps,

 \begin{figure}[!t]
    \centering
    \includegraphics[width=0.48\textwidth, height=0.32\textheight]{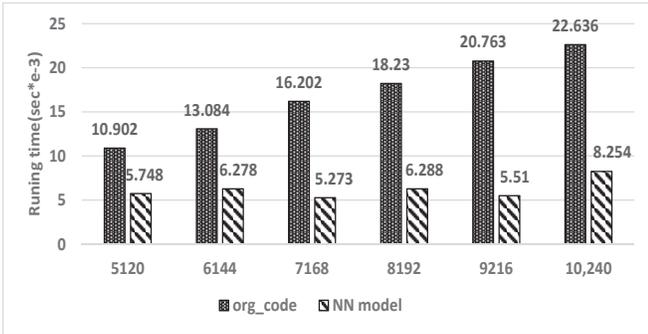}
       \vspace{-35pt}
        \caption{The execution times of the original Newton-Raphson method and our NN model.} 
  \vspace{-5pt}
  \label{fig:Newton_time_compare}
\end{figure}

Figure~\ref{fig:a small model} shows and compares the execution times of the Newton-Raphson method and the NN model solving a number of equations (from 5120 to 10240 equations). 
In general, the NN model has better performance than the Newton-Raphson method. We have up to 2.7x performance speedup. 
We also notice that as the number of equations increases, the execution time of the NN increases slowly,  while the execution time of the Newton-Raphson method increases almost linearly. Intuitively, the execution time of the neural network should increase linearly. We attribute such a slow increase in the execution time to the possible internal parallelism in the machine learning framework (particularly TensorFlow) we use to run the model.

\textbf{Discussion.}
The execution time (or the number of iterations to converge) of the Newton-Raphson method has a strong correlation to the initial guess of the solution (i.e., $x0$).
If the initial guess is close to the final solution, the equation can be solved quickly with the Newton-Raphson method.
For such case, it is difficult to use a neural network to perform better than the Newton-Raphson method (in terms of execution time).
However, finding a good initial guess of the solution is challenging, especially for a high-order equation.
We expect that in most cases, using a neural network to replace the Newton-Raphson method is promising. 

In our study, we choose a random data as the initial guess for the Newton-Raphson method and the neural network. Both of them use the same initial guess.
Also, to counter the potential effect of the initial guess on the execution time, we solve a number of equations, each of which uses different initial guess. We measure the execution time of solving all of the equations instead of solving the individual equations, shown in Figure~\ref{fig:a small model}.



\subsection{The L-J potential in LAMMPS}

The L-J potential is the most time-consuming computation in our study (we use in.lj.5 as the input problem of LAMMPS).  Table~\ref{tab:lj_pro} shows the profiling results of LAMMPS. The L-J potential takes more than 90\% of the total execution time. 


\begin{table}[]
\centering
\scriptsize
\caption{The profiling result of the L-J potential simulation.}
\label{tab:lj_pro}
\begin{tabular}{|p{1.1cm}|p{1.4cm}|p{1.4cm}|p{1cm}|p{1cm}|}
\hline
Section   &min time(sec) &avg time(sec)   &max time(sec)      &total   \\
\hline   \hline
Pair(the LJ potential)   & 12.609 &12.609 &12.609  &92.58\% \\ 
\hline 
Neigh  & 0.93452 &0.93452 & 0.93452 &6.86\%  \\
\hline
Comm   & 0.033068 & 0.033068  &0.033068  &0.24\%  \\
\hline
Output & 0.00012207 &0.00012207 & 0.00012207 & 0.00\% \\
\hline
Modify  & 0.034445 &0.034445 &0.034445  &0.25\% \\
\hline
Others  &  &0.008681 &  &0.06\% \\
\hline
\end{tabular}
\end{table}

Algorithm~\ref{alg:ForceCalculation} shows the major computation of the L-J potential. 
The algorithm involves a two-level loop: the outer loop uses the iterator $i$, and the inner loop uses the iterator $j$.
The inner loop traverses all neighbors $j$ of the atom $i$, calculates forces, and accumulates each force to the total force of atom $i$ (Line 6 in Algorithm~\ref{alg:ForceCalculation}.).

\begin{algorithm}[h]
\caption{\small The L-J Force calculation}
\label{alg:ForceCalculation}
\begin{algorithmic}[1]
\REQUIRE {Computing range $start$, $end$; cutoff distance $r_{cut}$; location array $atom$.}
\ENSURE {Force array $F$.}
\FOR{$i$ ranges from $start$ to $end$}
    \FOR{each neighbors $j$ of $i$ }
        \STATE $d_{ij}^2 \leftarrow |atomLocation[i]-atomLocation[j]|^2$;
        \IF {$d_{ij}^2 < r_{cut}^2$}
            \STATE calculate $Force_{ij}$;
            \STATE $TotalForce_{i}\leftarrow TotalForce_{i}+Force_{ij}$;
        \ENDIF
    \ENDFOR
    \STATE $F[i] \leftarrow TotalForce_{i}$;
\ENDFOR
\RETURN F

\end{algorithmic}
\end{algorithm}

We first consider replacing the inner loop with a neural network.
The inner loop calculates the interaction between atoms $i$ and $j$. 
As the following force calculation step is based on the neighbor cell space, it involves frequent calculation and memory access.
However, due to the irregular arrangement of molecular position, the numbers of atoms in different neighbor cells are different from each other.
Furthermore, impacted by the L-J force, atoms involving plenty of motions move frequently through a neighbor cell, which means the number of atoms in the same neighbor cell even change after several simulation time steps. 
The changed data size creates a major obstacle for model training and model reuse.
In the meanwhile, the average atoms in each neighbor cell are $265$ in our experiments, which requires a more sophisticated and elaborate model to learn motion patterns.

In the molecular simulation with the force of the LJ potential, interactions (repellent and attraction) usually take place at a pair of atoms.
Although a pair of atoms is almost the smallest unit during the L-J force calculation, most of these events are executed on atom pairs, illustrated at Line  $2-8$ in Algorithm~\ref{alg:ForceCalculation}. 
Taking the consideration of efficiency, the same-sized chunk of data enable to utilize the simplest neural network topology to offer the greatest rewards of QoR and model reuse.
Hence, we use a pair of atoms as our input data size. 
It is guaranteed that this same-sized chunk of data enables be read and write at the peer start and end points of procedure.

For the calculation of the L-J force, not only the force calculation but also the condition of distance (Line 4 in Algorithm~\ref{alg:ForceCalculation}) should be involved in the consideration of model design.
The reason is that the condition statement can prevent compiler optimization and efficient instruction scheduling. Removing the condition statement is beneficial for performance.

We use a simple 3-layers fully connected model (i.e., a $1\times3\times 1$ model) to replace the original code. We use an activation function which is a combination of ReLU and Tahn.


\begin{figure}[!t]
    \centering
    \includegraphics[width=0.50\textwidth, height=0.40\textheight]{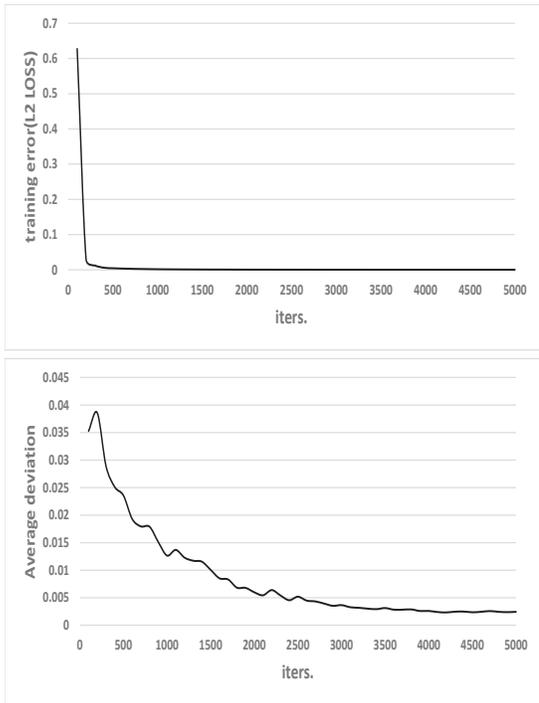}
       \vspace{-15pt}
        \caption{
        The training loss and the deviation for the training loss at each iteration.
        } 
  \vspace{-5pt}
\label{fig:LAMMPS_l2loss}
\end{figure}

After $5,000$ training steps, we achieve a $2.46$ speedup with an average derivation of $0.0026$ on the overall data. 
Figure~\ref{fig:LAMMPS_l2loss} illustrates how the training error (L2 loss) and test accuracy vary as we increase the number of training steps.
The figure shows that the training error and test accuracy converge after $5,000$ steps. 

\section{Evaluation (2pages)}
\label{sec:eval}

\subsection{Newton-Raphson method}





We use NNs of different topologies to 
replace the Newton-Raphson Method. 
We then study the accuracy and efficiency of the new Newton-Raphson method.
The accuracy of the new Newton-Raphson method using NN models of different topologies 
is presented in Table~\ref{tab:newton_err_vari}.
In the table, The first column is the topology for
NN models; 
the second column is the time spent 
in training; the third column
is the training step; the fourth column is the L2 loss;
the next column is the time spent in testing; the last column is the prediction accuracy; each row shows the result for a specific topology when
we use a NN of the specific topology 
to replace the Newton-Raphson method.

We can see that when a NN of a more complex topology is used, we spend 
more time on training and testing but we get a smaller
L2 loss in training and better accuracy in testing. In other words,
a more complex topology can help
decrease the L2 loss in training and 
increase the accuracy in testing, but 
doesn't cause overfitting. This means 
that a prediction accuracy of 95\% is
not our upper bound --- we can achieve
an even better prediction accuracy than 95\% when
we use a more complex topology than $3\times11\times8\times5\times1$ for NN. 

Furthermore, the best prediction accuracy we achieve is 73\% when a three-layered topology is used; the best prediction accuracy we achieve is 93\% when a four-layered topology is used; 
the best prediction accuracy we achieve is 95\% when a five-layered topology is used. 
This result suggests that the number 
of layers to the model topology to NN
has huge impact on the prediction accuracy. 
The more layers in the 
topology, the better prediction accuracy we can achieve. 
However, this benefit cannot sustain 
and decrease shortly when more layers 
are added to the topology. Moreover, 
for the same number of layers, 
more nodes in the hidden layer help 
increase the prediction accuracy. 
For example, for four-layered topologies, when the hidden layer is $3\times2$, 
the prediction accuracy is 69\%; the prediction accuracy is 89\% when the hidden
layer is $5\times3$; 
the prediction accuracy is 93\% when
the hidden layer is $8\times5$.

\begin{table}[]
\centering
\scriptsize
\caption{The variance of the prediction accuracy for Newton-Raphson method using different NN models.}
\label{tab:newton_err_vari}
\begin{tabular}{|p{1.1cm}|p{1.4cm}|p{0.8cm}|p{0.8cm}|p{1.4cm}|p{1cm}|}
\hline
Model topology      &Training time(sec*e) &training steps  &L2 Loss      &Running time(sec)           & Average prediction accuracy   \\
\hline   \hline
$3\times3\times1$    & 66.50 &5,000 &0.056 &4,407 & 51\% \\ 
\hline 
$3\times5\times1$   & 59.33 &5,000 &0.032 &5,190 & 54\% \\
\hline
$3\times8\times1$   & 61.14 &5,000 &0.032 &4,700 & 73\% \\
\hline
$3\times3\times2\times1$ & 69.95 &5,000 &0.040 &5,852 & 69\%\\
\hline
$3\times5\times3\times1$  & 66.18 &5,000 &0.032 &5,922 & 89\%\\
\hline
$3\times8\times5\times1$   & 75.87 &5,000 &0.026 &6,579 & 93\%\\
\hline
$3\times5\times3\times2\times1$ & 98.36 &5,000 &0.039 &8,931 & 89\%\\
\hline
$3\times8\times5\times3\times1$ & 94.45 &5,000 &0.035 &7,751 & 94\%\\
\hline
$3\times11\times8\times5\times1$ & 82.38 &5,000 &0.031 &7,453 & 95\%\\
\hline
\end{tabular}
\end{table}

\subsection{The L-J potential in Lammps}
\begin{table}[]
\centering
\scriptsize
\caption{The variance of the prediction accuracy for the L-J potential in Lammps using different NN models. }
\label{tab:lammps_err_vari}
\begin{tabular}{|p{0.8cm}|p{0.9cm}|p{0.7cm}|p{0.85cm}|p{0.8cm}|p{1.2cm}|p{0.85cm}|}
\hline
Model topology  &Training time(sec) &training steps  &L2 Loss(e-5) &Learning rate      &Running time(sec*e-6)  & Average absolute error   \\
\hline   \hline
$1\times3\times1$ & 48.51 &10,000 &7.14 & 0.005         & 5118           & 0.00261                  \\ %
\hline 
$1\times5\times1$ & 49.92 &10,000 &21.85 & 0.005        & 3970            & 0.0036                   \\ %
\hline
$1\times8\times1$ & 48.90 &10,000 &11.49 & 0.005        & 5085            &  0.0023                   \\  
\hline
$1\times3\times2\times1$ & 60.58 &10,000 &9.12 & 0.01        & 6585           & 0.0028                   \\
\hline
$1\times3\times5\times1$ & 60.64 &10,000 &12.28 & 0.01       & 7522           & 0.0022                    \\
\hline
$1\times5\times8\times1$ & 66.27 &10,000 &12.15 & 0.01      & 7104           & 0.0015                    \\
\hline
\end{tabular}
\end{table}
Similar to what we perform to the 
Newton-Raphson method, we replace the L-J 
potential in LAMMPS with NNs of various
topologies. We present the result for prediction
accuracy in Table~\ref{tab:lammps_err_vari}.
Note that we cannot calculate the
prediction error in this experiment
because some of the ground-truth values
are zero; we cannot calculate the relative error when the denominator is zero. Thus, we use the absolute error,
which is counted by the difference between the predicted value and the 
ground truth, as the metric for accuracy.

From the results, we can make the same conclusion that the addition of more neurons and layers leads to better prediction accuracy. The best prediction accuracy (0.0015 for the absolute error) is achieved by using the NN with the topology of $1\times5\times8\times1$.

\begin{figure}[!t]
    \centering
    \includegraphics[width=0.50\textwidth, height=0.35\textheight]{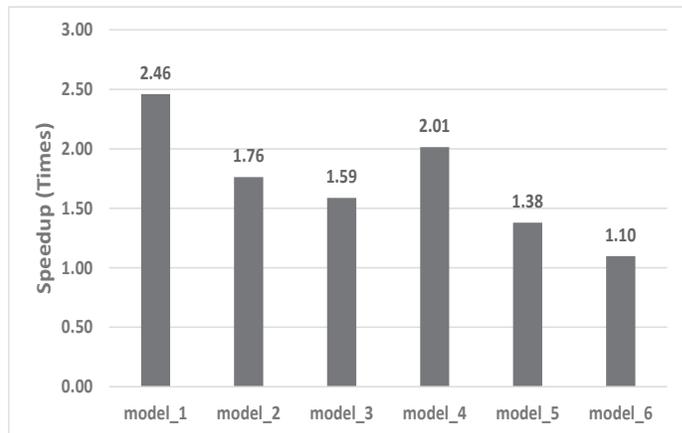}
       \vspace{-45pt}
	    \caption{Speedup achieved by NN models.} 
  \vspace{-15pt}
\label{fig:Lammps_time_compare}
\end{figure}

Figure~\ref{fig:Lammps_time_compare} presents the result for the efficiency study. We show the speedup achieved by using NN models in Table~\ref{tab:newton_err_vari} compared to the original execution as the baseline. 
We achieve 1.7X speedups on average by
using different topologies for NN. 
The variation in speedup depends on the complexity of model topology.
We can see a trend that 
the more nodes and more layers a NN has, the less speedup we can achieve.

\section{Related Work (1 page)}
\label{sec:related_work}

Machine learning has shown preliminary success when it is applied in HPC recently. 
We classify how machine learning is used in HPC into the following three cases.

\subheader{Enhancement Methods.} 
Machine learning has been used to enhance and
augment scientific applications to analyze very large data sets to reveal properties that are too complex to be discovered by previous systems. 
From predicting Molecular energetics to  tracking neutrinos, machine learning-driven 
enhancement dramatically advances the efficiency and accuracy of the solution
of well-known scientific problems~\cite{computation.llnl.gov}.
There are couples of simulation examples that successfully apply 
enhancement methods to their research field.

In weather prediction, Racah et al.~\cite{nips17:racha} use a semi-supervised multichannel spatiotemporal CNN 
model to realize a better localization of extreme weather. 
In cancer treatment, U.S. DOE laboratories, as well as the National Institutes of Health (NIH), have recently
launched a synthetic project, CANDLE~\cite{candle.cels}, targeting the top challenges in cancer diagnosis and treatment. At the current stage, researchers are leveraging information of millions of cancer patient records to diagnose cancer and figure out
the best treatment strategy using a 
scalable DNN for modeling. 

Moreover, in particle physics, 
George and Huerta~\cite{george2018deep} use
GPUs to accelerate training DNN for 
fast detection and 
processing gravitational wave data; the new 
machine learning-based approach is 
much efficient and resilient to noise than established gravitational wave detection algorithms. 
Seismologists and geophysicists recently reveal that machine learning techniques
can help them identify earthquake patterns from three years of earthquake records at The Geysers in California, one of the world's oldest and largest geothermal reservoirs~\cite{holtzman2018machine}. This is an unprecedented achievement. The subtle difference
between patterns is unseen by traditional
methods, which are less accurate. 
The patterns help researchers find
the fluctuating amounts of water injected belowground during the energy-extraction process.

\subheader{Modulation Methods.}
Besides that, machine learning has also be used to create refined input data for the next iteration round in a scientific simulation to modulate the simulation process. There are several successful examples. B. Wigley et al.~\cite{sr15:widley} propose a machine learning-based online optimization process for the production of Bose-Einstein condensates (BEC). With the repeated machine learning led learning, the optimization process finds
the optimal evaporation ramp for BEC production shortly in fewer iterations.
In Thermal–hydraulic modeling~\cite{carli2015incorporating}, an NN is trained using the output from simulations and then used to learn the dynamic behavior of the heated line. The NN is then added to the 4C circuit model as a new part. 
The NN model enables online control 
and fast assessment of the dynamic thermal-hydraulic system.
Similarly, Richard et al.~\cite{richard2014artificial} apply
an NN to ITER magnets aiming 
to predict when a disruption will occur in order to avoid damage to ITER and to adjust the reaction to keep generating power. The NN-based approach exceeds the best traditional methods in accuracy (95\% v.s. 85\%). 

\subheader{Approximation Methods.}
Approximation Methods becomes a favorite field in the last two years.
Our work, in essence, belongs to approximation methods. Approximation methods can be leveraged to shorten execution or save energy
by trading computation accuracy.
Approximation Methods use machine learning approximation to replace scientific simulation 
 Those code replacements happen at a coarse granularity. Typically the whole scientific simulation (instead of the fine-grained code regions) is replaced. 

There are a couple of successful cases, such as using machine learning to reproduce molecular energy surfaces~\cite{ml_fft} and simulate infrared spectra for molecular dynamics~\cite{md_ml17}.
In~\cite{ml_fft}, researchers use 
a DNN to replace Discrete Fourier Transform (DFT). By doing so, the Quantum chemistry (QC) simulation achieves
10e4x speedup with a high accuracy. 
After that, developing new drugs can be accomplished in minutes that would
have taken more than 10 years. 
Similarly, in~\cite{md_ml17}, 
an NN is used to reproduce the potential energy surface (PES) of a chemical system using the data computed by quantum chemistry methods. NN potentials can realize the accuracy of the underlying quantum chemical method, but also can be several-orders-of-magnitude faster using only several hundreds of electronic structure points.

Machine learning approximation is also managed to be used to speed up
quantum computing kernels~\cite{carleo2017solving}, in 
which Carleo and Troyer apply machine learning approximation on one of greatest challenges in quantum physics: the many-body problem, which describes
the complex correlations within the 
many-body wave function. Carleo and Troyer use an NN to reproduce the quantum many-body wave function. This forces the neural network to learn properties of the ground state of the wave function. This machine learning-based approach outperforms the state-of-the-art numerical simulation methods in accuracy.


In Computer Science, Approximation methods has been explored
in many sub-fields, including hardware~\cite{Esmaeilzadeh:2012:ASD:2150976.2151008, 6569370, Ranjan:2014:ASA:2616606.2617119, Sampson:2013:ASS:2540708.2540712}, compilers~\cite{baek2010green, Misailovic:2010:QSP:1806799.1806808, Sampson2015ACCEPTAP, sidiroglou2011managing}, programming languages~\cite{5523464, Rinard:2006:PAB:1183401.1183447, samadi2013sage, Sampson:2011:EAD:1993498.1993518}, and runtime systems~\cite{Campanoni:2015:HRP:2738600.2738630, Goiri:2015:ABA:2694344.2694351, Hoffmann:2011:DKR:1950365.1950390}. 
Approximate methods have been applied to 
many applications, such as streaming applications~\cite{khudia2015rumba, Krishnan:2016:IDA:2872427.2883026, asplos14:samadi},
However, there are only a few cases in  HPC applications (e.g., molecular dynamics simulation~\cite{carleo2017solving}, atmospheric modeling~\cite{Duben20130276}
and large-scale eigen decomposition~\cite{Zhang:2015:AAC:2840819.2840934}). We want to test
more HPC applications to extend 
approximation methods in HPC.
\section{Conclusion}
\label{sec:con}
Neural networks have gained prominence in recent years and we deploy them on approximate computing to chase for better performance.
This work motivates and introduces the machine learning-based approximate approach to mimic and replace original code regions.
We find that the potential code regions may gain best rewards from transformation with similar intrinsic characteristics. 
Based on these insights, we follow these guidelines for selecting a target code region to replace and designing a corresponding NN model. 
We implement two applications, the Newton-Raphson method, and Lennard-Jones (LJ) potential in LAMMPS, to realize our assumption.
As the result clearly show, NN models accelerate the original code region without introducing a huge deviation.
Various NN models provide appreciable speedup and accurate data depends on the data types, model complexity, and model reusability. 


\bibliographystyle{IEEEtran}
\begin{spacing}{0.9}
\bibliography{li}

\begin{thebibliography}{10}
\providecommand{\url}[1]{#1}
\csname url@samestyle\endcsname
\providecommand{\newblock}{\relax}
\providecommand{\bibinfo}[2]{#2}
\providecommand{\BIBentrySTDinterwordspacing}{\spaceskip=0pt\relax}
\providecommand{\BIBentryALTinterwordstretchfactor}{4}
\providecommand{\BIBentryALTinterwordspacing}{\spaceskip=\fontdimen2\font plus
\BIBentryALTinterwordstretchfactor\fontdimen3\font minus
  \fontdimen4\font\relax}
\providecommand{\BIBforeignlanguage}[2]{{%
\expandafter\ifx\csname l@#1\endcsname\relax
\typeout{** WARNING: IEEEtran.bst: No hyphenation pattern has been}%
\typeout{** loaded for the language `#1'. Using the pattern for}%
\typeout{** the default language instead.}%
\else
\language=\csname l@#1\endcsname
\fi
#2}}
\providecommand{\BIBdecl}{\relax}
\BIBdecl

\bibitem{ichep16:radovic}
A.~Radovic, ``{Neutrino Identification with a Convolutional Neural Network in
  the NOvA Detectors},'' in \emph{International Conference on High Energy
  Physics}, 2016.

\bibitem{sr15:widley}
P.~B.~Wigley, P.~J.~Everitt, A.~Hengel, J.~Bastian, M.~A.~Sooriyabandara,
  G.~McDonald, K.~Hardman, C.~D.~Quinlivan, M.~Perumbil, C.~c. Noschang~kuhn,
  I.~R.~Petersen, A.~Luiten, J.~Hope, N.~Robins, and M.~Hush, ``Fast
  machine-learning online optimization of ultra-cold-atom experiments,''
  vol.~6, 07 2015.

\bibitem{nips17:racha}
E.~Racah, C.~Beckham, T.~Maharaj, S.~Kahou, M.~Prabhat, and C.~Pal,
  ``{ExtremeWeather: A Large-scale Climate Dataset for Semi-supervised
  Detection, Localization, and Understanding of Extreme Weather Events},'' in
  \emph{NIPS}, 2017.

\bibitem{candle_anl}
{Argonne National Lab}, ``{{CANDLE: Exascale Deep Learning and Simulation
  Enabled Precision Medicine for Cancer}},'' http://candle.cels.anl.gov.

\bibitem{sidiroglou2011managing}
S.~Sidiroglou-Douskos, S.~Misailovic, H.~Hoffmann, and M.~Rinard, ``Managing
  performance vs. accuracy trade-offs with loop perforation,'' in
  \emph{Proceedings of the 19th ACM SIGSOFT symposium and the 13th European
  conference on Foundations of software engineering}, 2011.

\bibitem{5470469}
J.~Mengte, A.~Raghunathan, S.~Chakradhar, and S.~Byna, ``Exploiting the
  forgiving nature of applications for scalable parallel execution,'' in
  \emph{IEEE International Symposium on Parallel Distributed Processing
  (IPDPS)}, 2010.

\bibitem{Rinard:2006:PAB:1183401.1183447}
M.~Rinard, ``Probabilistic accuracy bounds for fault-tolerant computations that
  discard tasks,'' in \emph{PInternational Conference on Supercomputing}, 2006.

\bibitem{Rinard:2007:UEP:1297027.1297055}
M.~C. Rinard, ``Using early phase termination to eliminate load imbalances at
  barrier synchronization points,'' in \emph{Proceedings of the 22Nd Annual ACM
  SIGPLAN Conference on Object-oriented Programming Systems and Applications},
  2007.

\bibitem{Campanoni:2015:HRP:2738600.2738630}
S.~Campanoni, G.~Holloway, G.-Y. Wei, and D.~Brooks, ``Helix-up: Relaxing
  program semantics to unleash parallelization,'' in \emph{IEEE/ACM
  International Symposium on Code Generation and Optimization}, 2015.

\bibitem{plimpton2007lammps}
S.~Plimpton, P.~Crozier, and A.~Thompson, ``Lammps-large-scale atomic/molecular
  massively parallel simulator,'' \emph{Sandia National Laboratories}, vol.~18,
  p.~43, 2007.

\bibitem{micro2012neural}
H.~Esmaeilzadeh, A.~Sampson, L.~Ceze, and D.~Burger, ``Neural acceleration for
  general-purpose approximate programs,'' in \emph{Proceedings of the 2012 45th
  Annual IEEE/ACM International Symposium on Microarchitecture (Micro)}, 2012.

\bibitem{Esmaeilzadeh:2012:ASD:2150976.2151008}
------, ``Architecture support for disciplined approximate programming,'' in
  \emph{ASPLOS}, 2012.

\bibitem{carleo2017solving}
G.~Carleo and M.~Troyer, ``Solving the quantum many-body problem with
  artificial neural networks,'' \emph{Science}, vol. 355, no. 6325, pp.
  602--606, 2017.

\bibitem{ml_fft}
J.~S. Smith, O.~Isayev, and A.~E. Roitberg, ``Ani-1: an extensible neural
  network potential with dft accuracy at force field computational cost,''
  2017.

\bibitem{md_ml17}
M.~Gastegger, J.~Behlerb, and P.~Marquetand, ``{Machine Learning Molecular
  Dynamics for the Simulation of Infrared Spectra },'' \emph{Chemical Science},
  pp. 6695--7270, 2017.

\bibitem{khudia2015rumba}
D.~S. Khudia, B.~Zamirai, M.~Samadi, and S.~Mahlke, ``Rumba: An online quality
  management system for approximate computing,'' in \emph{Computer Architecture
  (ISCA), 2015 ACM/IEEE 42nd Annual International Symposium on}, 2015.

\bibitem{Krishnan:2016:IDA:2872427.2883026}
D.~R. Krishnan, D.~L. Quoc, P.~Bhatotia, C.~Fetzer, and R.~Rodrigues,
  ``Incapprox: A data analytics system for incremental approximate computing,''
  in \emph{WWW}, 2016.

\bibitem{asplos14:samadi}
M.~Samadi, D.~A. Jamshidi, J.~Lee, and S.~Mahlke, ``{Paraprox: Pattern-Based
  Approximation for Data Parallel Applications},'' in \emph{Architectural
  Support for Programming Languages and Operating Systems (ASPLOS)}, 2014.

\bibitem{Zhang:2015:AAC:2840819.2840934}
Q.~Zhang, Y.~Tian, T.~Wang, F.~Yuan, and Q.~Xu, ``Approxeigen: An approximate
  computing technique for large-scale eigen-decomposition,'' in \emph{ICCAD},
  2015.

\bibitem{cesm_models}
``The community climate system model,''
  \texttt{http://www.cesm.ucar.edu/models/ccsm4.0/}.

\bibitem{computation.llnl.gov}
``A converging path for simulation, machine learning, and big data,''
  \url{https://computation.llnl.gov/newsroom/simulation-machine-learning-big-data-converging-path},
  accessed APRIL 18, 2016.

\bibitem{candle.cels}
R.~Stevens, ``Exascale deep learning and simulation enabled precision medicine
  for cancer,'' \url{http://candle.cels.anl.gov/}, argonne National Laboratory.

\bibitem{george2018deep}
D.~George and E.~Huerta, ``Deep learning for real-time gravitational wave
  detection and parameter estimation: Results with advanced ligo data,''
  \emph{Physics Letters B}, vol. 778, pp. 64--70, 2018.

\bibitem{holtzman2018machine}
B.~K. Holtzman, A.~Pat{\'e}, J.~Paisley, F.~Waldhauser, and D.~Repetto,
  ``Machine learning reveals cyclic changes in seismic source spectra in
  geysers geothermal field,'' \emph{Science advances}, vol.~4, no.~5, p.
  eaao2929, 2018.

\bibitem{carli2015incorporating}
S.~Carli, R.~Bonifetto, L.~Savoldi, and R.~Zanino, ``Incorporating artificial
  neural networks in the dynamic thermal--hydraulic model of a controlled
  cryogenic circuit,'' \emph{Cryogenics}, vol.~70, pp. 9--20, 2015.

\bibitem{richard2014artificial}
L.~S. Richard, R.~Bonifetto, S.~Carli, A.~Froio, A.~Foussat, and R.~Zanino,
  ``Artificial neural network (ann) modeling of the pulsed heat load during
  iter cs magnet operation,'' \emph{Cryogenics}, vol.~63, pp. 231--240, 2014.

\bibitem{6569370}
J.~Han and M.~Orshansky, ``Approximate computing: An emerging paradigm for
  energy-efficient design,'' in \emph{ETS}, 2013.

\bibitem{Ranjan:2014:ASA:2616606.2617119}
A.~Ranjan, A.~Raha, S.~Venkataramani, K.~Roy, and A.~Raghunathan, ``Aslan:
  Synthesis of approximate sequential circuits,'' in \emph{DATE}, 2014.

\bibitem{Sampson:2013:ASS:2540708.2540712}
A.~Sampson, J.~Nelson, K.~Strauss, and L.~Ceze, ``Approximate storage in
  solid-state memories,'' in \emph{ACM TOCS}, 2014.

\bibitem{baek2010green}
W.~Baek and T.~M. Chilimbi, ``Green: a framework for supporting
  energy-conscious programming using controlled approximation,'' in \emph{ACM
  Sigplan Notices}, vol.~45, no.~6.\hskip 1em plus 0.5em minus 0.4em\relax ACM,
  2010, pp. 198--209.

\bibitem{Misailovic:2010:QSP:1806799.1806808}
S.~Misailovic, S.~Sidiroglou, H.~Hoffmann, and M.~Rinard, ``Quality of service
  profiling,'' in \emph{ICSE}, 2010.

\bibitem{Sampson2015ACCEPTAP}
A.~Sampson, A.~Baixo, B.~Ransford, T.~Moreau, J.~Yip, L.~Ceze, and M.~Oskin,
  ``Accept: A programmer-guided compiler framework for practical approximate
  computing,'' \emph{University of Washington Technical Report UW-CSE-15-01},
  2015.

\bibitem{5523464}
V.~K. Chippa, D.~Mohapatra, A.~Raghunathan, K.~Roy, and S.~T. Chakradhar,
  ``Scalable effort hardware design: Exploiting algorithmic resilience for
  energy efficiency,'' in \emph{Design Automation Conference}, 2010.

\bibitem{samadi2013sage}
M.~Samadi, J.~Lee, D.~A. Jamshidi, A.~Hormati, and S.~Mahlke, ``Sage:
  Self-tuning approximation for graphics engines,'' in \emph{Microarchitecture
  (MICRO), 2013 46th Annual IEEE/ACM International Symposium on}, 2013.

\bibitem{Sampson:2011:EAD:1993498.1993518}
A.~Sampson, W.~Dietl, E.~Fortuna, D.~Gnanapragasam, L.~Ceze, and D.~Grossman,
  ``Enerj: Approximate data types for safe and general low-power computation,''
  in \emph{PLDI}, 2011.

\bibitem{Goiri:2015:ABA:2694344.2694351}
I.~Goiri, R.~Bianchini, S.~Nagarakatte, and T.~D. Nguyen, ``Approxhadoop:
  Bringing approximations to mapreduce frameworks,'' in \emph{ASPLOS}, 2015.

\bibitem{Hoffmann:2011:DKR:1950365.1950390}
H.~Hoffmann, S.~Sidiroglou, M.~Carbin, S.~Misailovic, A.~Agarwal, and
  M.~Rinard, ``Dynamic knobs for responsive power-aware computing,'' in
  \emph{ASPLOS}, 2011.

\bibitem{Duben20130276}
P.~D. D{\"u}ben, J.~Joven, A.~Lingamneni, H.~McNamara, G.~De~Micheli, K.~V.
  Palem, and T.~N. Palmer, ``On the use of inexact, pruned hardware in
  atmospheric modelling,'' \emph{Philosophical Transactions of the Royal
  Society of London A: Mathematical, Physical and Engineering Sciences}, vol.
  372, no. 2018, 2014.

\end{thebibliography}
\end{spacing}
\clearpage


\end{document}